\documentclass[reqno]{amsart}
\usepackage{amsmath, amsfonts, amsthm}
\newtheorem{thm}{Theorem}

\theoremstyle{remark}
\newtheorem*{remark}{Remark}

\newcommand{\N}{{\mathbb N}}

\newcommand{\Z}{{\mathbb Z}}

\title[Horizon Spectrum]{Spectral Geometry of Cosmological and Event Horizons for Kerr-Newman de Sitter metrics\\}
\author{Martin Engman}
\author{Gerardo A. Santana}
\begin{document}
\address{Departamento de Ciencias y Tecnolog\'{i}a, Universidad Metropolitana, 
San Juan, PR 00928}
\email{um\_mengman@suagm.edu}
\email{mathengman@yahoo.com}
\subjclass[2000]{Primary 58J50; Secondary 83C15, 83C57}
\begin{abstract}
{We study the Laplace spectra of the intrinsic instantaneous metrics on the event and cosmological horizons of a 
Kerr-Newman de Sitter space-time and prove that the spectral data from these horizons uniquely determine the space-time. 
This is accomplished by exhibiting formulae relating the parameters of the space-time metric to the traces of 
invariant and equivariant Green's operators associated with these Laplacians. In particular, an interesting explicit 
formula for the 
cosmological constant is found.} 
\end{abstract}
\maketitle

\section{Introduction}
The Kerr-Newman de Sitter metric exhibits four horizons. Three of these have physical interpretations as: a Cauchy (inner)
 horizon, 
a black-hole event horizon, and a 
cosmological horizon. In this paper, we study the two largest: the event and cosmological horizons. 

The space-time metric induces on each of these horizons an $S^1$-invariant Riemannian metric 
which we 
call {\em intrinsic instantaneous metrics}. Each of these metrics' Green's operators yield $S^1$-invariant and 
$k$-equivariant trace formulae. In \cite{engri} we proved, for the case of a single horizon in the Kerr metric, that 
the two trace formulae uniquely determine the two parameters of this space-time. In the present paper there are four 
parameters which must be
 determined:  the mass, angular momentum, charge, and cosmological constant. It is, therefore, 
fortunate that there are two horizons
 each 
providing two distinct types of trace formulae which produce a non-linear system of four equations in the parameters 
which can be shown to 
have a unique solution. Each of the parameters is given semi-explicitly in terms of the spectrum but in the case of the 
cosmological constant we are, in fact, able to find an explicit formula.

\section{ The metric}

The {\em Kerr-Newman de Sitter metric}  (see, for example \cite{gibhawk}) is a solution of the Einstein equations 
with positive cosmological constant $\Lambda$ given by:

\begin{equation}
ds^2=-\frac{\Delta_{r}}{\chi^2\rho^2}(dt-a\sin^2\theta d\phi)^2+\frac{\Delta_{\theta}\sin^2\theta}{\chi^2\rho^2}
\left[adt-(r^2+a^2)d\phi\right]^2+\rho^2\left(\frac{dr^2}{\Delta_{r}}+\frac{d\theta^2}{\Delta_{\theta}}\right) 
\label{eq:knds}
\end{equation}\\
where 
$\Delta_{\theta}=1+\frac{1}{3}\Lambda a^2\cos^2\theta$, $\chi = 1+ \frac{1}{3}\Lambda a^2$, 
$\Delta_{r}=(r^2+a^2)(1-\frac{1}{3}\Lambda r^2)-2mr+Q^2$ and $\rho = r^2 + a^2 \cos^2 \theta$. 
The parameters $(m, a, Q, \Lambda)$ (all of which, unless otherwise stated, we assume to be positive) represent respectively, the total mass, angular momentum per unit mass, 
the charge, and the cosmological constant.  
The uncharged 
($Q=0$) case produces the {\em Kerr de Sitter metric}.
 
 In general, of course, the equation $\Delta_r = 0$ has four roots. Here we are only interested in two positive real 
roots, denoted by
 $r_e$ and $r_{c}$, which correspond to the event and cosmological horizons respectively. We assume, of course, that  $r_e < r_{c}$.
  
  Let $r_0$ denote either $r_e$ or $r_c$. To obtain the intrinsic instantaneous metrics on these surfaces we pull 
back the space-time metric 
\eqref{eq:knds} to the surface defined by $r=r_0$
 (so that $dr=0$) and $dt=0$ to obtain two dimensional Riemannian metrics on each of the event
 and cosmological horizons. Both take the form: 

\begin{equation}
ds^2=\frac{\rho^2}{\Delta_{\theta}}d\theta^2+\frac{\Delta_{\theta}(r_0^2+a^2)^2}{\chi^2\rho^2}\sin^2\theta d\phi^2
\label{eq:surfkerr}
\end{equation}

Following the notation of \cite{smarr},we define {\em the scale parameter} by $\eta=\sqrt{r_0^2+a^2}$ and 
{\em the distortion parameter} by 
$\beta = \frac{a}{\sqrt{r_0^2+a^2}}$. We also define a new parameter 
\begin{equation}
\xi=\frac{\Lambda \frac{a^2}{3}}{1+\Lambda \frac{a^2}{3}}  \label{eq:xi}
\end{equation}
with the change of variable  $x=-\cos \theta$ one finds that the horizon metric is 

\begin{equation}
 ds_{r_0}^2=\eta^2(1-\xi)\left(\frac{1}{f(x)}dx^2 + f(x)d\phi^2\right)  \label{eq:xeh}
\end{equation} 
where $(x,\phi) \in(-1,1) \times [0,2\pi)$ and 
\begin{equation} 
f(x)=\frac{1-\xi(1-x^2)}{1-\beta^2(1-x^2)}(1-x^2). \label{eq:function}
\end{equation} 

The area of this metric is $A=4\pi \eta^2(1-\xi)$ (see \cite{davies}).
It is well known that the Gauss curvature of such a metric takes the form $K(x)=-f''(x)/(2\eta^2(1-\xi))$ so that in
 this 
case from \eqref{eq:function}
 the curvature is:

\begin{equation}
K(x)=\frac{1}{\eta^2(1-\xi)}\left[\frac{\xi}{\beta^2}+\left(1-\frac{\xi}{\beta^2}\right)\frac{1-\beta^2(1+3x^2)}
{(1-\beta^2(1-x^2))^3}\right] \label{eq:curv}
\end{equation}\\
The special case $\beta^2 = \xi$  gives a constant curvature metric on a horizon, but yields only one horizon 
corresponding to a positive 
solution of $\Delta_r=0$.

\section{Spectrum of $S^1$ invariant metrics}

For any Riemannian manifold with metric $g_{ij}$ {\em the Laplacian}, in local coordinates, is given by

$$\Delta_g =- \frac{1}{\sqrt{g}} \frac{\partial }{\partial x^i} \left( \sqrt{g}g^{ij} 
\frac{\partial }{\partial x^j} \right) . $$

This is the Riemannian version of the {\em Klein-Gordon}, or {\em D'Alembertian}, or 
{\em wave} operator usually denoted by $\Box$. 

In this section we outline some our previous work on the spectrum of the Laplacian on $S^1$ invariant metrics on $S^2$. 
The interested reader may consult \cite{eng1}, \cite{eng2}, 
\cite{eng3}, and \cite{zeld} for further details. 

To simplify the discussion the area of the metric is normalized to $A=4\pi$ for this section only.
The metrics we study have the form:
\begin{equation}
dl^2=\frac{1}{f(x)}dx^2 + f(x)d\phi^2  \label{eq:surfrev}
\end{equation}
where $(x,\phi) \in(-1,1) \times [0,2\pi)$ and $f(x)$ satisfies $f(-1)=0=f(1)$ and 
$f'(-1)=2=-f'(1)$. 
In this form, it is easy to see that the 
Gauss curvature of this metric is given by 
$K(x) =(-1/2)f^{''}(x)$.
 The canonical (i.e. constant curvature) metric is obtained by taking 
$f(x)=1-x^{2}$ and the metric \eqref{eq:xeh} 
is a homothety (scaling) of a particular example of the general form 
\eqref{eq:surfrev}. 

The Laplacian for the metric \eqref{eq:surfrev} is

$$\Delta_{dl^2} =  -\frac{\partial }{\partial x} \left( f(x) 
\frac{\partial }{\partial x} \right) - \frac{1}{f(x)}\frac{\partial^2 }{\partial \phi^2}.$$
Let $\lambda$ be any eigenvalue of $-\Delta$.
We will use the symbols $E_{\lambda}$ and $\dim E_{\lambda}$ to denote the 
eigenspace for $\lambda $ and it's multiplicity (degeneracy) respectively. In this paper 
the symbol $\lambda_m$ will always mean the $m$th \underline{distinct} 
eigenvalue. We adopt the convention $\lambda_0=0$. 
Since $S^1$ (parametrized here by $0 \leq \phi <2\pi $) acts
 on $(M,g)$ by isometries we can separate variables and because $\dim E_{\lambda_m} \leq 2m+1$ (see 
\cite{eng3} for the proof), the 
orthogonal decomposition of $E_{\lambda_m}$ has the special form

\begin{equation*}
E_{\lambda_m}= \bigoplus_{k=-m}^{k=m} e^{ik\phi}W_k
\end{equation*}
in which $W_k (=W_{-k})$ is the ``eigenspace" (it might contain only $0$)
 of the ordinary differential operator

\begin{equation*}
L_{k}=-\frac{d}{dx}\left(f(x)\frac{d}{dx}\right) + \frac{k^2}{f(x)}
\end{equation*}
with suitable boundary conditions. It should be observed that $\dim W_k \leq 1$, a value of zero for this dimension 
occuring when $\lambda_m$ is not in the spectrum of
$L_k$.

The set of positive eigenvalues is given by $Spec(dl^2)= \bigcup_{k\in \Z} Spec L_k$ 
and consequently the nonzero part of the
spectrum of $-\Delta$ can be studied via the spectra 
$Spec L_{k}=\{0<\lambda_{k}^{1} < \lambda_{k}^{2} < \cdots 
<\lambda_{k}^{j} < \cdots \} \forall k \in \Z$. 
The eigenvalues $\lambda^j_0$ in the case $k=0$ above are called the 
{\em $S^1$ invariant eigenvalues} since their eigenfunctions are 
invariant under the action of the $S^1$ isometry group. If $k \neq 0$ the eigenvalues
are called {\em $k$ equivariant} or simply {\em of type $k \neq 0
$}. Each $L_{k}$ has a Green's operator, 
$\Gamma_{k}:(H^{0}(M))^{\perp} \rightarrow L^{2}(M)$, whose spectrum 
is $\{ 1/\lambda_{k}^{j} \}_{j=1}^{\infty}$, and whose trace is defined by 
\begin{equation}
\gamma_{k} \equiv \sum_{j=1}^{\infty} \frac{1}{\lambda_{k}^{j}}. \label{eq:trace}
\end{equation}

The formulas of present interest were derived in \cite{eng1} and \cite{eng2} and are given by  

\begin{equation} \gamma_{0}=
   \frac{1}{2} \int_{-1}^{1} \frac{1-x^{2}}{f(x)} dx \label{eq:gam_0}
\end{equation}
and
\begin{equation}  
   \gamma_{k}=\frac{1}{|k|} \hspace{.2in}  \mbox{if $k \neq 0$} \label{eq:gamk}
\end{equation}

\begin{remark}

 One must be careful with the definition of $\gamma_0$ since 
$\lambda^0_0 = 0$ is an $S^1$ invariant eigenvalue of $-\Delta$. To avoid this difficulty we studied the $S^1$ 
invariant 
spectrum
of the Laplacian on 1-forms in \cite{eng2} and then observed that the 
nonzero eigenvalues are the same for functions and 1-forms.

\end{remark}

\section{Spectral Determination of Horizons}

In case $f(x)$ is given by \eqref{eq:function} the metric \eqref{eq:xeh} is related to \eqref{eq:surfrev} via the
 homothety
$ds_{r_0}^2=\eta^2(1- \xi)dl^2$, and it is well known that 

$$\lambda \in Spec(dl^2) \hspace{.2in} \mbox{if and only if} \hspace{.2in} \frac{\lambda}{\eta^2(1-\xi)} 
\in Spec(\eta^2(1-\xi) dl^2)$$ 
so that, after an elementary integration and some algebra, the trace formulae (\eqref{eq:gam_0} and \eqref{eq:gamk}), 
for either horizon, take the form

\begin{equation}
 \gamma_{0}=\eta^2\left[1-\beta^2+\left(\xi-\beta^2\right)\frac{\left(
\sqrt{\frac{1-\xi}{\xi}}\arctan\sqrt{\frac{\xi}{1-\xi}}\right)-1}{\xi}\right] 
\label{eq:ehgam0}
\end{equation}
and
 \begin{equation}  
   \gamma_{k}=\frac{\eta^2(1-\xi)}{|k|} \hspace{.2in}  \forall k \neq 0. \label{eq:ehgamk}
\end{equation}
An immediate consequence of \eqref{eq:ehgamk} is that the area of the metric has a representation for each 
$k \in \N$ given
by
\begin{equation}
A=4\pi k\gamma_k. \label{eq:area}
\end{equation}

We now define:
\begin{equation} g(\xi) =\frac{\left(
\sqrt{\frac{1-\xi}{\xi}}\arctan\sqrt{\frac{\xi}{1-\xi}}\right)-1}{\xi}.\label{eq:g}
\end{equation}

Then equation \eqref{eq:ehgam0} has the form:
\begin{equation}
\label{eq:gam0}
\gamma_{0}=\eta^2\left[1-\beta^2+\left(\xi-\beta^2\right)g(\xi)\right].
\end{equation}

One now has, for each of the two physical horizons $r=r_e$ and $r=r_c$, corresponding parameters and traces denoted
 by $\eta_e$, $\beta_e$, $\gamma_k^e$ and $\eta_c$, $\beta_c$, $\gamma_k^c$ respectively.

Using the definitions  of the parameters and the system of equations consisting of \eqref{eq:ehgamk} and
 \eqref{eq:gam0} (for $k=1$) for both horizons, one easily obtains the following formula for the cosmological constant.

\begin{thm}
If $a\neq 0$, $\Lambda >0$, and  $r_c\neq r_e$ then 
\begin{equation}
\label{eq:cosmconst}
\Lambda = \frac{3(\gamma_0^e-\gamma_1^e+\gamma_1^c-\gamma_0^c)}{\gamma_1^c\gamma_0^e-\gamma_1^e\gamma_0^c}.
\end{equation}
\qed
\end{thm} 

After defining
\begin{equation}
\label{eq:hxi}
h(\xi)=\frac{1+\xi g(\xi)}{1-\xi},
\end{equation}
the pairs of equations  \eqref{eq:ehgamk} and \eqref{eq:gam0} for each horizon
yield 
\begin{equation}
\label{eq:hxispec}
h(\xi)=\frac{\gamma_0^e-\gamma_0^c}{\gamma_1^e-\gamma_1^c}.
\end{equation} And once it is  verified that \eqref{eq:hxi} is invertible, we obtain
\begin{equation}
\label{eq:xispec}
\xi = h^{-1}\left(\frac{\gamma_0^e-\gamma_0^c}{\gamma_1^e-\gamma_1^c}\right).
\end{equation}
From the definition \eqref{eq:xi}, the angular momentum parameter is given by:
\begin{equation}
\label{eq:asq}
a^2= \frac{h^{-1}\left(\frac{\gamma_0^e-\gamma_0^c}{\gamma_1^e-\gamma_1^c}\right)}
{1-h^{-1}\left(\frac{\gamma_0^e-\gamma_0^c}{\gamma_1^e-\gamma_1^c}\right)}\cdot
 \frac{\gamma_1^c\gamma_0^e-\gamma_1^e\gamma_0^c}{(\gamma_0^e-\gamma_1^e+\gamma_1^c-\gamma_0^c)}.
\end{equation}

From \eqref{eq:ehgamk} ($k=1$), \eqref{eq:xispec}, and \eqref{eq:asq} one can solve for $r_e$ and $r_c$ respectively. 
The resulting equations are:

\begin{equation}
\label{eq:rec}
r^2_{e,c}= \frac{\gamma_1^{e,c}}{1-h^{-1}\left(\frac{\gamma_0^e-\gamma_0^c}
{\gamma_1^e-\gamma_1^c}\right)} - \frac{h^{-1}\left(\frac{\gamma_0^e-\gamma_0^c}{\gamma_1^e-\gamma_1^c}\right)}
{1-h^{-1}\left(\frac{\gamma_0^e-\gamma_0^c}{\gamma_1^e-\gamma_1^c}\right)}\cdot 
\frac{\gamma_1^c\gamma_0^e-\gamma_1^e\gamma_0^c}{(\gamma_0^e-\gamma_1^e+\gamma_1^c-\gamma_0^c)}
\end{equation}
and we have $\Lambda$, $a^2$, $r_e$ and $r_c$ in terms of the traces. Finally, after substituting these (distinct) 
values of $r$ into the equation  $\Delta_r =0$, a nonsingular linear system in the variables $m$ and $Q^2$ is obtained 
and, therefore, $m$ and $Q^2$ are uniquely determined.

We have thus proved:
\begin{thm}
The Kerr-Newman de Sitter space-time is uniquely determined by the union of the 
spectra of the cosmological and
 event horizons.\qed
\end{thm}

\section{Discussion}

The spectra we study in this paper should not be confused with the quasinormal mode frequencies arising from the angular 
part of the 
Teukolsky master equation (see \cite{bat}, \cite{fn}, \cite{t}, \cite{n}, \cite{suz} and references therein, among many 
others).
On the other hand, in certain limiting cases for the parameters, they do coincide but we will not pursue this comparison 
here.  

The reader may have noticed that Theorem 2 is consistent with
 the holographic principle (\cite{sus}, \cite{th}) in as much as the structure of the ($3+1$ dimensional) Kerr-Newman 
de-Sitter
 space-time is encoded in the intrinsic spectral data of the two (two-dimensional) horizon surfaces.

Many of the calculations in this paper are independent of which pair of horizons are being used. This 
leads to the conjecture that one can 
use the spectra of the inner Cauchy horizon, together with that of the event horizon to obtain the uniqueness result 
and the formula for $\Lambda$. 

\section{Acknowledgements} 
 A special thanks goes out to Mar\'{i}a del Rio for her support, especially for the second author, during the writing 
of this paper. 
This work was partially supported by the NSF Grants: Model Institutes for Excellence and AGMUS Institute of 
Mathematics at UMET.

\end{document}